\RequirePackage{fix-cm}

\documentclass[smallextended]{svjour3}

\usepackage{amsmath}  
\usepackage{amsfonts} 
\usepackage[american]{babel}

\newcommand{\ud}{\text d}

\DeclareMathOperator{\re}{Re}
\DeclareMathOperator{\im}{Im}

\newcommand{\rstate}{| I\rangle} 
\newcommand{\ahat}{\hat A}
\newcommand{\abar}{\bar A}
\newcommand{\bhat}{\hat B}
\newcommand{\mhat}{\hat M}
\newcommand{\phat}{\hat p}
\newcommand{\xhat}{\hat x}

\journalname{Foundations of Physics}

\begin{document}

\title{Anomalous weak values are caused by disturbance}

\author{Asger C.~Ipsen}
\institute{Asger C.~Ipsen \at Farum, Denmark
			\email{acipsen@gmail.com}}

\date{\today}

\maketitle

\begin{abstract}
In combination with post-selection, weak measurements can
lead to surprising results known as anomalous weak values.
These  lie outside the bounds of the spectrum of the relevant observable,
as in the canonical example
of measuring the spin of an electron (along some axis) to be 100.
We argue that the
disturbance caused by the weak measurement, while small, is
sufficient to significantly affect the measurement result,
and that this is the most reasonable explanation of anomalous
weak values.
\end{abstract}

\section{Introduction}

The subject we want to address can be somewhat confusing 
at the conceptual level. So, before getting to the actual arguments, we will
start with a rather long introduction where we try to explain the main
issues as clearly as possible.

For concreteness we will discuss weak values in the context they
where originally introduced\cite{AAV88,AV90}, namely that of a
weakly interacting limit of a von Neumann measurement.
Let us begin by recalling the standard von Neumann model.
It describes the measurement of an observable $\ahat$ as a two-step process;  first the
system of interest, initially in state $|\psi\rangle$, interacts with the
measurement apparatus (or meter) according to
\begin{equation}
  |\psi\rangle \rstate \to
  e^{-i\lambda\ahat\phat}|\psi\rangle\rstate .
  \label{eq:vonN}
\end{equation}
The meter is modeled by a single canonical degree of freedom $[\xhat,\phat] = i$
and, for the purposes of this paper, we will take the initial state $\rstate$ to be
a Gaussian with zero mean and unit variance,
\begin{equation}
  \langle x  \rstate := \sqrt{G(x)},\qquad
    G(x) := (2\pi)^{-1/2}e^{-x^2/2} .
  \label{eq:gauss}
\end{equation}
The parameter $\lambda\in\mathbb{R}$ controls the strength of the interaction.
Note that we will always assume that the Hamiltonian of the isolated system 
vanishes.
The procedure is completed by a sharp measurement of $\xhat$, resulting in $x$ with
probability (we assume that the system is finite dimensional)
\begin{align}
  P_\lambda(x|\psi) &= \sum_i \left|\langle a_i | \langle x|
  e^{-i\lambda\ahat\phat}|\psi\rangle \rstate\right|^2 \nonumber\\
  &= \sum_i |\langle a_i|\psi\rangle|^2|\langle x| e^{-i\lambda a_i\phat}\rstate|^2\nonumber\\
  &= \sum_i |\langle a_i|\psi\rangle|^2 G(x-\lambda a_i)
  \label{eq:x-dist-no-post-strong}
\end{align}
For sufficiently large $\lambda$ the different peaks of the distribution can be
resolved, and a result $x\simeq \lambda a_i$ corresponds to a specific eigenvalue of $\ahat$.

In the opposite limit of small $\lambda$ the peaks corresponding to different eigenvalues
combine to a single peak centered on the expectation value of $\ahat$,
\begin{equation}
  P_\lambda(x|\psi) = G(x-\lambda \langle\ahat\rangle_\psi) + O(\lambda^2),\qquad
  \langle\ahat\rangle_\psi := \langle\psi|\ahat|\psi\rangle .
  \label{eq:x-dist-no-post}
\end{equation}
We will use the term \emph{weak measurement} 
to mean that $\lambda$ is sufficiently small that $O(\lambda^2)$ terms can be neglected.
To obtain the \emph{weak value} we extend the measurement procedure just described 
by performing a \emph{post-selection} on the system after the von Neumann interaction
\eqref{eq:vonN}. That is, 
we perform a projective measurement with $|\varphi\rangle$ as a one of the eigenstates,
and only keep the measurement outcomes for the runs
of the experiment where the result of the final measurement is $|\varphi\rangle$.
The resulting distribution of $x$ conditioned on successful post-selection is
\begin{align}
  P_\lambda(x|\varphi,\psi) &= P_\lambda(\varphi|\psi)^{-1}\left|\langle \varphi | \langle x|
  e^{-i\lambda\ahat\phat}|\psi\rangle \rstate\right|^2\nonumber\\
   &= G(x-\lambda \re \ahat_w) + O(\lambda^2),
  \label{eq:x-dist-pps}
\end{align}
with the weak value $\ahat_w$ defined by
\begin{equation}
  \ahat_w := \frac{\langle\varphi|\ahat|\psi\rangle}{\langle\varphi|\psi\rangle} .
  \label{eq:wv}
\end{equation}
The normalization factor is the unconditional post-selection probability
$P_\lambda(\varphi|\psi) = |\langle\varphi|\psi\rangle|^2 + O(\lambda)$.
We have assumed that $|\varphi\rangle$ is not orthogonal to $|\psi\rangle$, 
and we will continue to do so for the remainder of the paper.

The purpose of the interaction \eqref{eq:vonN} is to displace the meter observable $\xhat$
proportionally to the system observable $\ahat$, but it will in
general also cause some uncontrolled  change in the state of the system itself. 
This is the back-action or disturbance of the measurement. 
The parameter $\lambda$ directly controls the strength of the interaction, and so,
for weak measurements where $\lambda$ is small, the amount of disturbance is
also small. 
The central question we are interested in is whether it is small enough to be neglected.
More precisely, we want to investigate the hypothesis of\\
\textbf{Disturbance Insignificance (DI)}:
\emph{To linear order in $\lambda$
the disturbance (of the system) caused by the intermediate weak measurement can be neglected.
In particular, the order $\lambda$ shift of the conditional $x$ distribution 
\eqref{eq:x-dist-pps}
is not affected by the disturbance.}\\
An alternative formulation of DI is given in Appendix \ref{app:state-update}, and
a more general hypothesis is discussed in Appendix \ref{app:lindblad}.

A Fourier transformation of \eqref{eq:gauss} shows that the distribution of $p$
is Gaussian with zero mean and a variance of $1/4$. 
Substituting a normal distributed random number $p$ for $\phat$
as a heuristic model, we 
would then expect the back-action
of \eqref{eq:vonN}, in a given run of the experiment, and for small $\lambda$, to be roughly
\begin{equation}
  \Delta|\psi\rangle \sim -i \lambda p \ahat|\psi\rangle .
\end{equation}
This is of the same order of magnitude as the shift of the meter (i.e.~$\lambda$),
so  it is not \emph{a priori} clear
that DI is a good assumption.
The main claim
of the present paper is that DI does \emph{not} follow from
ordinary quantum mechanics, but that it rather is an additional postulate which
leads to surprising or paradoxical results in some situations.

A central concept in the weak value literature is that of a pre- and post-selected
(PPS) ensemble. Given a number of identical systems
 all prepared in the state 
$|\psi\rangle$ at time $t_1$, consider the subensemble of systems found to be in
state $|\varphi\rangle$ when we make a projective measurement
on each system at a later time $t_2$. This subensemble is said to be the PPS ensemble
defined by states $|\psi\rangle$ and $|\varphi\rangle$.
Note that no other interaction with the systems is to take place between $t_1$ 
and $t_2$.
This ensemble is unusual in that, as soon as we learn which systems belong to it, i.e.~at $t_2$,
the systems are strongly and irreversibly perturbed by the projective measurement.
We cannot make an ordinary measurement at an intermediate time $t$, $t_1 < t < t_2$,
 since doing so would influence
the post-selection. However, \emph{if} we assume DI, we should be able to perform
a weak measurement of $\ahat$ on each system
without significantly altering the ensemble.
We are then naturally lead to consider the average of the measurement results 
corresponding
to post-selected systems as a kind of expectation value of $\ahat$ in the 
PPS ensemble.\footnote{For example,  it is stated that
  ``[\ldots] we can operationally interpret the \emph{real part} of the weak value [\ldots] 
    as  an \emph{idealized conditioned average of $\ahat$ in the limit of zero
    measurement disturbance}.'' in Ref.~\cite{DJ12A}.}
This average is just the conditional expectation value which,
according to \eqref{eq:x-dist-pps}, is (proportional to) the real part of $\ahat_w$.

It is important to note that the understanding of the weak value as the 
result of a measurement of $\ahat$ in a PPS
ensemble necessarily rests on DI. Without DI the weak measurement
 must  be considered  an active part of the process that
selects the members of the ensemble.
But then it is not reasonable to think of the PPS ensemble as an
independently existing thing on which we perform some measurement.\footnote
  {In the case of strong intermediate measurements, a similar objection 
  regarding post-selected ensembles was raised
  in Ref.~\cite{BB86}.}
For this reason we claim that, while a weak measurement by itself constitute a
legitimate, if imprecise, measurement, \emph{the combination of a weak measurement
and post-selection should not be considered as any kind of measurement 
of $\ahat$.}\footnote{A post-selected weak measurement is certainly a legitimate
  experimental procedure which leads to the conditional expectation value
  $\re\ahat_w$. However, without the assumption of DI, it is not reasonable to
  interpret this value as the result of a measurement of $\ahat$.}

The issue might by illustrated 
by a simple example\cite{AAV88}. Suppose we, instead of post-selecting on the system, 
decide to only keep data from runs with, say,  $x \geq 100\lambda$. Then the conditional
distribution becomes
\begin{equation}
  P(x|``x \geq 100\lambda",\psi) =
  \begin{cases}
    \mathcal{N}G(x-\lambda \langle\ahat\rangle_\psi) + O(\lambda^2), &x \geq 100\lambda \\
    0, &x<  100\lambda 
  \end{cases},
\end{equation}
with $\mathcal{N}$ a normalization constant. Clearly the expectation value of $x$
would be larger than $100\lambda$ in this case, but no one  would
interpret this to mean that
we have constructed an ensemble where the expectation value of $\ahat$ is larger than $100$.
It would also not be reasonable to claim that this
procedure constitutes a measurement of $\ahat$ in any 
sense, since we are clearly changing the result by ``cherry picking'' data points.
This example is so different from the actual protocol for obtaining the weak value that
the comparison might seem silly. However, note that the interaction \eqref{eq:vonN}
perturbs the system in a way which depends on the state of the measurement apparatus.
This means that, by post-selecting on the system, one is also indirectly post-selecting
on the meter.
So post-selecting on a observable of the system might have some similarities with
our artificial example of post-selecting on the position of the meter after all.

Returning to weak values, the crucial point is that,
since the post-selection occurs after the weak measurement,
the interaction between system and measurement apparatus can directly influence
the probability of successful post-selection. 
This can lead to unintuitive statistical effects.
When $\lambda$ is made very small, so
as to disturb the system very little, the amount of information contained in the data
of each experimental run also becomes very small. 
Our basic conclusion can then be phrased as:
The statistical amplification (by combining data from many runs of the experiment)
necessary to determine the weak value is also sufficient to amplify these small unintuitive
effects caused by the disturbance of the intermediate weak measurement.
A number published works contain statements that appear to contradict our findings.
We will not attempt an exhaustive review of the literature, but a representative
list of examples is Refs.~\cite{Vai96,RLS04,Tol07C,Hof10,DJ12A,BFB13,VBDKWMSBW17,Coh17},
see also Appendix \ref{app:quotes}.
It should be noted that a few authors have, by different arguments, reached conclusions
similar to ours\cite{Hu90,FC14,Kas17}.

A final notion we need to introduce is that of an \emph{anomalous weak value}. The ordinary
expectation value $\langle\ahat\rangle$ is bounded between the smallest and
largest eigenvalue of $\ahat$. On the other hand, it is always
possible\footnote{\label{foot:anomalous}
  As long as
  $\ahat$ is not proportional to the identity, one can find
  states $|\psi\rangle$, $|\bot\rangle$ such 
  that $\langle\bot|\psi\rangle = 0$ and
  $\langle\bot|\ahat|\psi\rangle > 0$. With 
  $|\varphi\rangle = \sqrt{1-\epsilon^2}|\bot\rangle+\epsilon|\psi\rangle$ 
  and $\epsilon\in [-1,1]\setminus \{0\}$
  sufficiently small, $\re \ahat_w = \epsilon^{-1}\langle\bot|\ahat|\psi\rangle + O(\epsilon^0)$
  is anomalous.}
to find states $|\psi\rangle$, $|\varphi\rangle$ such that $\re \ahat_w$ is outside
these bounds. When this happens we say that the weak value is anomalous.
This is a striking and unintuitive phenomenon which 
was even the basis for the title of the first paper on weak values\cite{AAV88}.

From the point of view that the weak value is a kind of expectation value it is 
difficult to accommodate such anomalous values.
Indeed, suppose we have $N$ copies of the system prepared in the state
 $|\psi\rangle$ and we perform an ordinary 
(strong) measurement of $\ahat$ on each. Each measurement  results in
some eigenvalue $a_n$, $n=1,\ldots,N$.
One could now argue
that there should be some subensemble $\mathcal{S} \subseteq \{1,\ldots,N\}$
(of size $|\mathcal{S}| \sim N|\langle \varphi|\psi\rangle|^2$) consisting of those systems
which would have been post-selected in state $|\varphi\rangle$,
\emph{had we not performed the intermediate measurement}.
In other words, the members
of $\mathcal{S}$ constitute a realization of the PPS ensemble. 
The expectation value
of $\ahat$ in $\mathcal{S}$ could then be given its usual meaning as the empirical mean
\begin{equation}
  \lim_{N\to\infty}\frac{1}{|\mathcal{S}|}\sum_{n\in\mathcal{S}}a_n ,
  \label{eq:average}
\end{equation}
but this quantity is clearly bounded like the ordinary expectation value,
and thus cannot in general be identified with $\re \ahat_w$.
Note that it is not possible to compute the average \eqref{eq:average} in an 
actual experiment,
since measuring $\ahat$ (strongly) makes it impossible to determine $\mathcal S$.
Now, if one, as we are advocating, understands the shift in the distribution of $x$ as
partly due to post-selection effects, the phenomenon of anomalous weak values becomes
a non-mystery. Indeed, there appears to be no reason that such statistical effects
should be bounded
by the spectrum of $\ahat$. In the next section we will see in an explicit example
that disturbance and post-selection can in fact lead to this kind of unbounded shift.

The remainder of the paper is structured as follows:
Section \ref{sec:significance} contains the main argument.
We first demonstrate that, in a certain sense,
the mathematical structure of quantum mechanics allows the disturbance caused by
a weak interaction to appear smaller than it really is.
Then we show that the amount
of disturbance caused by a weak measurement is sufficient, 
in combination with post-selection, to cause shifts in the
conditional $x$ distribution of magnitude $\lambda$ 
(i.e.~of the same order as the shift in Eq.~\eqref{eq:x-dist-pps}).
From considerations of sequential weak measurements, an independent argument against 
DI is given in Section \ref{sec:sequential}.
Appendix \ref{app:state-update} contains some remarks on the notion of measurement
disturbance and a reformulation of DI, while Appendix \ref{app:quotes} 
quotes some passages from the literature that seem to endorse the 
hypothesis of DI.
Appendix \ref{app:collective} extends the reasoning of Section \ref{sec:significance}
to the case where  an entire ensemble is measured by  a single
measurement apparatus, and Appendix \ref{app:lindblad} discusses 
some related more technical work due
to Dressel and Jordan.

\section{The significance of a little disturbance}
\label{sec:significance}

We begin by considering the unconditional probability 
$P_\lambda(\varphi|\psi)$ of successful post-selection more carefully.
It is clear that any deviation from its unperturbed value $|\langle\varphi|\psi\rangle|^2$
must be due to back-action from the intermediate measurement. Expanding in the coupling
 $\lambda$ we find
\begin{align}
  P_\lambda(\varphi|\psi) &:= \int\ud x \left|\langle \varphi |\langle x|
  e^{-i\lambda\ahat\phat}|\psi\rangle\rstate\right|^2\nonumber\\
   &= \int\ud x\, G(x)\left|\langle\varphi|\left(1+\lambda\ahat\frac x 2
   + \frac{\lambda^2}{2} \ahat^2\left[\frac{x^2}{4}-\frac 1 2\right]
   +O(\lambda^3)\right)|\psi\rangle\right|^2\nonumber\\
  &= |\langle \varphi |\psi\rangle|^2\left[1+\frac{\lambda^2}{4}
  		\left(|\ahat_w|^2-\re[(\ahat^2)_w]\right)+O(\lambda^4)\right],
  \label{eq:postselection-prob}
\end{align}
where 
\begin{equation}
  (\ahat^2)_w := \frac{\langle\varphi|\ahat^2|\psi\rangle}{\langle\varphi|\psi\rangle} .
\end{equation}
The leading correction is of order $\lambda^2$, 
which suggests that back-action is a second order effect.
It thus appears as if the disturbance really is insignificant in the $\lambda\to 0$
limit.\footnote{Variants of this argument appears several places, 
e.g.~\cite{Tol07C,Hof10,BFB13,Coh17}.}
Despite the intuitive appeal of this reasoning, it is in fact misleading, as we will
now argue.

Consider what would happen if we, instead of the weak measurement, simply
gave the system a small unitary kick
\begin{equation}
  |\psi\rangle \to
  e^{-i\lambda\ahat x'/2}|\psi\rangle
  \label{eq:kick}
\end{equation}
with $x'$ drawn at random from the standard Gaussian distribution $G(x')$.
The post-selection probability is seen to be
\begin{align}
  P'_\lambda(\varphi|\psi) 
  &:= \int\ud x'\, G(x') 
    \left|\langle\varphi|e^{-i\lambda\ahat x'/2}|\psi\rangle\right|^2\nonumber\\
   &= |\langle \varphi |\psi\rangle|^2\left[1+\frac{\lambda^2}{4}
  		\left(|\ahat_w|^2-\re[(\ahat^2)_w]\right)+O(\lambda^4)\right]
  \label{eq:kick-postselection-prob}
\end{align}
exactly as in \eqref{eq:postselection-prob}. We see that
the disturbance is able to ``hide''; in each run of the experiment the system
received a kick of magnitude $\lambda$, but,
after marginalizing over $x'$ to get the unconditional probability
\eqref{eq:kick-postselection-prob}, the linear term drops out.

We can take this example  further by considering the conditional distribution
of $x'$,
\begin{align}
  P'(x'|\varphi,\psi) &:= 
     P'_\lambda(\varphi|\psi)^{-1}G(x') 
      \left|\langle\varphi|e^{-i\lambda\ahat x'/2}|\psi\rangle\right|^2\nonumber\\
   &= G(x'-\lambda \im \ahat_w) + O(\lambda^2) .
   \label{eq:x-dist-kick}
\end{align}
Notice the similarity to \eqref{eq:x-dist-pps}, even though no measurement of $\ahat$
is taking place. 
The distribution \eqref{eq:x-dist-kick} shows very explicitly that the disturbance
\eqref{eq:kick} \emph{is} significant, since it is entirely responsible for the shift.
This also means that we \emph{cannot} conclude from the smallness of the
correction found in Eq.~\eqref{eq:postselection-prob}
that DI holds.
The shift $\im\ahat_w$, like $\re\ahat_w$, can be made
arbitrarily large by appropriate choice of $|\psi\rangle$ and 
$|\varphi\rangle$.\footnote{This follows by the argument of Footnote \ref{foot:anomalous},
  but with $|\varphi\rangle = i\sqrt{1-\epsilon^2}|\bot\rangle+\epsilon|\psi\rangle$.}
As the shift is clearly a statistical effect in this case, it is not surprising that
it can exceed the spectrum of $\ahat$.

To make further contact between our `fake measurement' and an actual weak measurement,
it is useful to define the operator
\begin{equation}
  \xhat' := 2\phat .
\end{equation}
Our choice of $\rstate$ is such that the wavefunction in $\xhat'$-basis is 
identical to that in $\xhat$-basis,
\begin{equation}
  \langle x'  \rstate = \sqrt{G(x')} .
\end{equation}
We then see that $P'_\lambda(\varphi|\psi)$ can be expressed as
\begin{equation}
  P'_\lambda(\varphi|\psi) = \int\ud x' \left|\langle \varphi |\langle x'|
  e^{-i\lambda\ahat\phat}|\psi\rangle\rstate\right|^2 .
  \label{eq:kick-postselection-prob-alt}
\end{equation}
It is now clear that $P'_\lambda(\varphi|\psi) = P_\lambda(\varphi|\psi)$
for any $\lambda$, since the only difference between \eqref{eq:postselection-prob}
and \eqref{eq:kick-postselection-prob-alt} is whether we trace out the meter in
$\xhat$-basis or $\xhat'$-basis.
Similarly we find that $P'(x'|\varphi,\psi)$ is equal to the conditional probability
distribution for a measurement of $\xhat'$,
\begin{equation}
  P'_\lambda(x'|\varphi,\psi) = 
     P_\lambda(\varphi|\psi)^{-1}\left|\langle \varphi |\langle x'|
  e^{-i\lambda\ahat\phat}|\psi\rangle\rstate\right|^2 .
\end{equation}
The fact that the shift of $\phat \propto \xhat'$ is
proportional to $\im\ahat_w$ was already mentioned in Footnote 4 of \cite{AAV88}.

We now have two formulations of our fake weak measurement, both leading to the same probabilities.
One is in terms of a random unitary kick \eqref{eq:kick}, the other keeps the von Neumann
interaction \eqref{eq:vonN}, but we measure the meter observable $\xhat'$ instead
of $\xhat$. The reason that these lead to the same probabilities, is that $|x'\rangle$
is an eigenstate of $\exp(-i\lambda\ahat\phat)$ (since $\xhat' \propto \phat$), and that 
one can thus interchange the order of the interaction with the system and
the measurement of $\xhat'$.

Let us make a couple of further remarks:

1) The fact that the  distribution of a variable is shifted in some process involving 
post-selection does not mean that the underlying physical quantity was changed by the same
amount (on average).
In our example the value of $x'$ is chosen before the interaction with the system, and
it is never changed from this original value. Nevertheless, the conditional 
distribution is shifted
according to Eq.~\eqref{eq:x-dist-kick}. Returning to weak measurements, it follows that
we \emph{cannot} conclude from \eqref{eq:x-dist-pps} that the position of the meter is
shifted by $\lambda\re\ahat_w$ (on average).
We thus disagree with the claim of Ref.~\cite{AB05} that the
``weak value can be regarded as a
  \emph{definite} mechanical effect on a measuring probe\ldots''.

2) An interaction with the system which changes the post-selection 
probability to the one given in Eq.~\eqref{eq:postselection-prob}
is able to shift the distribution of $x$ an amount of order $\lambda$ purely
through post-selection effects. Indeed, take the interaction
\begin{equation}
  |\psi\rangle \rstate \to
  e^{-i\lambda\ahat\xhat/2}|\psi\rangle\rstate .
\end{equation}
By the above consideration, with $\xhat$ and $\xhat'$ interchanged, this leads to
the post-selection 
probability \eqref{eq:postselection-prob} and a conditional $x$ distribution
which is shifted by $\lambda\im \ahat_w$.

3) An interesting situation arises when the choice between measuring $\xhat$ and $\xhat'$ is
delayed to after post-selection of the system. If the choice falls on $\xhat'$ we have
seen that the disturbance must be considered significant. On the other hand, according
to DI, when measuring $\xhat$ we should be able to neglect the disturbance. But this 
would mean that, on assumption of DI, the experimenter is able to choose, \emph{after}
the post-selection
has taken place, whether said post-selection was significantly disturbed.
It is not at all clear how one is to make sense of this conclusion.

In summary we have failed to find any evidence for DI, i.e.~that the disturbance caused by the 
von Neumann interaction \eqref{eq:vonN} can be neglected. Instead it appears
natural to assume that this disturbance plays a significant role in the observed shift
in the conditional distribution of the meter observable $\xhat$. This is  also sufficient to
explain why the
weak value can exceed the spectrum of $\ahat$ under certain circumstances.

\section{Sequential weak measurements}
\label{sec:sequential}
An interesting situation where the presence of significant  back-action becomes apparent,
is the case of sequential weak measurements. That is, when several 
independent weak measurements are performed on the system between preparation and 
post-selection\cite{RS04,MJP07}.
For our purposes it is sufficient to consider two measurements. The setup
then consist of two identically prepared meters, and the combined
interaction corresponding to first measuring $\ahat$ then $\bhat$ is
\begin{equation}
  |\psi\rangle\rstate_1  \rstate_2 \to
  e^{-i\lambda_2\bhat\phat_2}e^{-i\lambda_1\ahat\phat_1}
  |\psi\rangle\rstate_1\rstate_2 .
  \label{eq:seq-interaction}
\end{equation}
The assumption of DI\footnote{I.e.~that the disturbance of the intermediate measurements is
  insignificant when neglecting $O(\lambda_1^2)$ and $O(\lambda_2^2)$ terms.} 
would imply that the two
measurements should not affect each other.
In particular, the order of the two unitaries of
\eqref{eq:seq-interaction} should be immaterial. 

The joint distribution, conditioned on post-selecting $|\varphi\rangle$, is found to be
\begin{align}
  &P_{\lambda_1\lambda_2}(x_1,x_2|\varphi,\psi)\nonumber\\ 
  &\qquad:=  
    P_{\lambda_1\lambda_2}(\varphi|\psi)^{-1}\left|\langle \varphi | \langle x_1|\langle x_2|
     e^{-i\lambda_2\bhat\phat_2}e^{-i\lambda_1\ahat\phat_1}
  |\psi\rangle\rstate_1\rstate_2\right|^2 \nonumber\\
  &\qquad= G(x_1-\lambda_1\re[\ahat_w])G(x_2-\lambda_2\re[\bhat_w])\nonumber\\
  &\qquad\qquad\times\left(1+\frac{\lambda_1\lambda_2}{2}\re[(\bhat\ahat)_w-\ahat_w\bhat_w]x_1x_2
               +O(\lambda_1^2)+O(\lambda_2^2)\right)
  \label{eq:x-dist-seq}
\end{align}
with
\begin{equation}
  (\bhat\ahat)_w := 
    \frac{\langle\varphi|\bhat\ahat|\psi\rangle}{\langle\varphi|\psi\rangle} .
\end{equation}
In addition to the expected shifts corresponding
to the weak values of $\ahat$ and $\bhat$ the last factor indicates that $x_1$ and $x_2$
have become correlated. This is not surprising, since the two measurements are performed on the
same system. What is more interesting is that this factor \emph{does} depend on the 
order of the measurements, contrary to the intuition from DI. 
Indeed, reversing the order corresponds to replacing 
$\re[(\bhat\ahat)_w] \to \re[(\ahat\bhat)_w]$ on the right hand side of \eqref{eq:x-dist-seq},
which in general makes a 
difference.\footnote{As long as $\ahat$ and $\bhat$ do not commute,
  $[\ahat,\bhat]$ is not proportional to the identity (we assume the system is finite
  dimensional). By the same argument as in
  Footnote \ref{foot:anomalous} we can find states $|\varphi\rangle,|\psi\rangle$
  such that $\re [([\ahat,\bhat])_w] \neq 0$. But then 
  $\re[(\bhat\ahat)_w] \neq \re[(\ahat\bhat)_w]$.} 
This phenomena seems very difficult to understand under the assumption
of DI.\footnote{In Ref.~\cite{BFB13} the fact that sequential weak measurements are sensitive to
  the ordering of the measurements is also discussed.
  But,
  since the authors have already decided according to other arguments that DI holds,
  they end up concluding that ``[w]eak measurements are then still 
  disturbing in some sense, although they do not disturb the state or later measurements''.}
  
As soon as one drops DI, it becomes clear that the 
order of the measurements should in general matter.
To see this it is illuminating to work out the distribution for a joint
measurement of $\xhat'_i = 2\phat_i$:
\begin{multline}
  P_{\lambda_1\lambda_2}(x'_1,x'_2|\varphi,\psi) =  
  G(x'_1-\lambda_1\im[\ahat_w])G(x'_2-\lambda_2\im[\bhat_w])\\
  \times\left(1+\frac{\lambda_1\lambda_2}{2}\re[\ahat_w\bhat_w-(\bhat\ahat)_w]x_1'x_2'
               +O(\lambda_1^2)+O(\lambda_2^2)\right)
  \label{eq:x-dist-seq-kick}
\end{multline}
We notice that the result is very similar to \eqref{eq:x-dist-seq}. In particular
this distribution will also depend on the order of the two interactions.
In this case, however, it is no mystery how this dependence comes about.
By the same reasoning as in Section \ref{sec:significance} we can view 
\eqref{eq:x-dist-seq-kick} as describing a process where we first draw two random
numbers $x_1',x_2'$, then perturb the system according to
\begin{align}
  |\psi\rangle &\to
  e^{-i\lambda_2\bhat x_2'/2}e^{-i\lambda_1\ahat x_1'/2}|\psi\rangle\nonumber \\
  &= \left(1-i\lambda_1\frac{\ahat}{2} x_1'-i\lambda_2 \frac{\bhat}{2} x_2'
     -\frac{\lambda_1\lambda_2}{2}\frac{\bhat\ahat}{2} x_1' x_2'
     +O(\lambda_1^2)+O(\lambda_2^2)\right)|\psi\rangle,
  \label{eq:seq-kick}
\end{align}
and finally post-select on $|\varphi\rangle$. It is clear that the term of order 
$\lambda_1\lambda_2$
reflects the fact that we are perturbing with $\ahat$ first and then with $\bhat$, and
that this is what is causing \eqref{eq:x-dist-seq-kick} to depend on the order of 
the unitaries.

\section{Conclusion}
Several phenomena connected with weak values  appear paradoxical
under the assumption of disturbance insignificance (DI). 
In particular we have discussed anomalous weak values and
the fact that the order matters when performing several weak measurements in sequence.
As we have argued in Section \ref{sec:significance} it is perfectly in accordance with
standard quantum mechanics to drop DI as a general principle.
Doing this allow us to understand these  `paradoxes' as mundane, 
albeit unintuitive, consequences 
of the ordinary laws of probability applied to experiments with weakly 
(but \emph{not} negligibly)
disturbing measurements and post-selection.

In this paper we have, for reasons of simplicity and concreteness, only considered weak measurement
of the von Neumann type.
The weak value can be obtained by 
much more general classes of measurement procedures. We refer the reader to
the very brief sketch in Appendix \ref{app:lindblad}, ~Refs.~\cite{DJ12C,DJ12B}
and references therein.
There is no type of weak measurement which less disturbing (on average) than
the von Neumann scheme\cite{Ips15}, so
our arguments and conclusions apply equally well to this broader class.

\begin{acknowledgements}
Isabell Lubanski Ipsen is thanked for helpful comments on the manu\-script, and
Josh Combes is thanked for inspiring discussion. Lev Vaidman is also thanked for
interesting correspondence.
\end{acknowledgements}

\appendix

\section{State update and disturbance}
\label{app:state-update}
For an ideal \emph{classical} measurement, it is usually assumed that the state of the
system after interacting with the apparatus is the same as it was before the interaction.
In contrast, the presence of measurement disturbance (or back-action) then means that this
condition is not satisfied, i.e.~that the meter somehow causes the system to change
state.

The situation for quantum measurements is more subtle. 
Since it is the system that will have our interest, it is helpful
to abstract away the `inner workings' of the measurement procedure.
As in the main text, let us concentrate on the von Neumann model.
We thus consider an apparatus that takes a quantum 
system as input 
 and produces a 
number $x$ and a quantum system as output. It can be completely characterized by two 
pieces of data.
One is the probability $P(x|\psi)$ of result $x$ given the input $|\psi\rangle$.
In our specific case $P(x|\psi) = P_\lambda(x|\psi)$ is given by 
Eq.~\eqref{eq:x-dist-no-post-strong}.
The second piece of data is the state of the
output $|\chi_{\psi,x}\rangle$ conditioned on a specific result $x$,
\begin{equation}
  |\chi_{\psi,x}^\lambda\rangle := 
    P_\lambda(x|\psi)^{-1/2}\langle x|e^{-i\lambda\ahat\phat}|\psi\rangle\rstate .
  \label{eq:cond-state}
\end{equation}

In analogy with the classical case, 
we want to define quantum measurement disturbance as the change of the
state of the system caused by the interaction with the 
meter (compare with e.g.~Refs.~\cite{Bus09,LS13}).
The question is how to compare the
state before and after interaction.

One approach would be to look at how close $|\chi_{\psi,x}^\lambda\rangle$ is to 
$|\psi\rangle$. Expanding \eqref{eq:cond-state} for small $\lambda$ we find
\begin{equation}
  |\chi_{\psi,x}^\lambda\rangle = 
    \left(1+\lambda[\ahat-\langle\ahat\rangle_\psi]\frac x 2 + O(\lambda^2)\right)|\psi\rangle .
  \label{eq:cond-state-first-order}
\end{equation}
From this point of view the conclusion would thus be that the the back-action is of order
$\lambda$ (unless $|\psi\rangle$ happens to be an eigenstate of $\ahat$), in conflict with
DI.

Seemingly, there is just one way to avoid this conclusion, which is to understand 
a pure quantum states as only expressing partial knowledge
about the system (similarly to how a probability distribution
expresses partial knowledge about a classical random variable).
Then it could be argued that the change $|\psi\rangle \to |\chi_{\psi,x}^\lambda\rangle$ is,
at least partly, a form of Bayesian update (or conditioning). 
That is, the claim would be that some of the
difference between $|\chi_{\psi,x}^\lambda\rangle$ and $|\psi\rangle$ reflects the fact
that we learn something
about the system by learning $x$, and should thus update our believes about the quantum 
system.
From this point of view we can reformulate DI as\\
\textbf{Disturbance Insignificance' (DI')}: \emph{Any disturbance caused by the measurement
process only shows up in the higher order (i.e.~$O(\lambda^2)$) correction to the conditional
state \eqref{eq:cond-state-first-order}. Hence the first order correction must be understood
as purely due to some form of Bayesian update.}\\
An interesting discussion of quantum Bayesian conditioning can be found in Section
V of Ref.~\cite{LS13}.

Let us  mention that it is not possible to interpret the 
update $|\psi\rangle \to |\chi_{\psi,x}^\lambda\rangle$  \emph{purely} as Bayesian
conditioning. That is, the update necessarily introduces some amount
of disturbance.
Indeed, if $|\psi\rangle \to |\chi_{\psi,x}^\lambda\rangle$ were a kind of Bayesian
update, then marginalizing over $x$ (i.e.~`unlearning' the measurement result) should
give back the original state, but we actually find\footnote{The map 
  $|\psi\rangle\langle\psi| \mapsto \int P_\lambda(x|\psi) 
        |\chi_{\psi,x}^\lambda\rangle\langle\chi_{\psi,x}^\lambda |\ud x$ 
  is sometimes called a non-selective update.}
\begin{equation}
  \rho_\psi^\lambda := \int P_\lambda(x|\psi) 
        |\chi_{\psi,x}^\lambda\rangle\langle\chi_{\psi,x}^\lambda |\ud x
    = \int G(x') e^{-i\lambda\ahat x'/2}|\psi\rangle
        \langle\psi|e^{i\lambda\ahat x'/2} \ud x'
\end{equation}
which is mixed (unless $|\psi\rangle$ is an eigenstate of $\ahat$),
and thus not equal to the initial state $|\psi\rangle\langle\psi|$. It is a general result that
\emph{any} quantum measurement must be disturbing in this 
sense\cite{Bus09,LS13}.\footnote{A precise statement is: If 
  $\rho_\psi = 
        |\psi\rangle\langle\psi|$ for all $\psi$, then the probability
  $P(x|\psi)$ is independent of $\psi$.}

The fact that the difference between $\rho_\psi^\lambda$ and $|\psi\rangle\langle\psi|$ is
solely due to back-action is used in Section \ref{sec:significance} to
quantify the amount of disturbance. Indeed, the difference between the overall post-selection
probability $P_\lambda(\varphi|\psi)$ (see Eq.~\eqref{eq:postselection-prob})
and its unperturbed value $|\langle\varphi|\psi\rangle|^2$ can be written as
\begin{equation}
  P_\lambda(\varphi|\psi) - |\langle\varphi|\psi\rangle|^2
    = \left\langle \rho_\psi^\lambda - |\psi\rangle\langle\psi| \right\rangle_\varphi .
\end{equation}

\section{Disturbance Insignificance in the literature}
\label{app:quotes}

We claim in the introduction that a number of works in the literature apparently conclude
or postulate
that the disturbance of weak measurements \emph{can} be neglected.
The subtle nature of the subject matter means that there is a substantial risk of
misunderstandings.
With that in mind we will quote some relevant passages from these papers.
After the quotes we make a couple of brief remarks.
\begin{itemize}
\item (Vaidman, 1996, p.~899)\cite{Vai96}:
  ``I propose to consider the standard measuring procedure [\ldots] in which we weaken
  the interaction in such a way that the state of the quantum system is not changed
  significantly during the interaction.''
\item (Resch, Lundeen \& Steinberg, 2004, p.~125)\cite{RLS04}:
  ``In particular, [the weak measurement strategy] makes it possible to contemplate the
  behavior of a system defined both by state preparation and by a later post-selection,
  without significant disturbance of the system in the intervening period.''
\item (Tollaksen, 2007, abstract)\cite{Tol07C}:
  ``When measurements are performed which do not disturb the pre- and post-selection
  (i.e.~weak measurements) [\ldots]''
\item (Tollaksen, 2007, p.~9063)\cite{Tol07C}:
  ``[\ldots] we have now shown that when considered as a limiting process, the disturbance 
  goes to zero more quickly than the shift in the measuring device, which means for a 
  large enough ensemble, information (e.g.~the expectation value) can be obtained even
  though not even a single particle is disturbed.''
\item (Hofmann, 2010, abstract)\cite{Hof10}:
  ``[\ldots] weak measurements have negligible back action [\ldots]''
\item (Hofmann, 2010, p.~2)\cite{Hof10}:
  ``For very small measurement strengths $\epsilon$, the effects of the quantum state
  on the measurement probabilities is linear in $\epsilon$ while the measurement 
  back action is quadratic in $\epsilon$. It is therefore possible to realize quantum
  state tomography with negligible back action.''
\item (Dressel \& Jordan, 2012, p.~7)\cite{DJ12A}:
  ``[the real part of the weak value] can be interpreted as an idealized limit point for
  the average of $\ahat$ in the initial state $\hat{\rho}_i$ that has been conditioned
  on the postselection $\hat{P}_f$ without any appreciable intermediate measurement 
  disturbance.''
\item (Bednorz, Franke \& Belzig, 2013, abstract)\cite{BFB13}:
  ``We show that it is possible to define general weak measurements, which are noninvasive:
  the disturbance becomes negligible as the measurement strength goes to zero.''
\item (Bednorz, Franke \& Belzig, 2013, p.~9)\cite{BFB13}:
  ``[\ldots] weak measurements (both classical and quantum) are noninvasive in a stronger
  sense: their disturbance vanishes as $g^2$ regardless of the type of measurements
  before/after.''
\item (Vaidman et al., 2017, p.~2)\cite{VBDKWMSBW17}:
  ``The change in the other systems [e.g.~a meter] should be large enough to be seen, but the back
  action on the system should be small enough, such that the change in the two-state vector
  describing the system can be neglected. Since we allow an unlimited ensemble of 
  experiments with identical pre- and postselection, the required limits are achievable.''
\item (Cohen, 2017, p.~1263)\cite{Coh17}: 
  ``[\ldots] for $\epsilon \ll 1$, which is indeed justified in the weak measurement
  regime, then the fidelity\footnote{
    The fidelity between the initial pure state $\psi$ and the state of the system after interaction
    with the meter is equal to the post-selection probability $P_\lambda(\psi|\psi)$. In the
    case of a von Neumann measurements we  find
    $1 - \lambda^2(\langle\ahat^2\rangle_\psi-[\langle\ahat\rangle_\psi]^2)/4$ by substituting
    $\varphi \to \psi$ in \eqref{eq:postselection-prob}.}
  is 1 up to $O(\epsilon^2)$. Therefore, by definition, the
  state has been negligibly disturbed, and this is now a precise claim.''
\item (Cohen, 2017, p.~1264)\cite{Coh17}: 
  ``[\ldots] weak measurements are non-invasive, that is, the probability of evolving
  the initial state to an orthogonal state through a weak measurement decreases like
  $g^2$. This property significantly limits the amount of backaction [\ldots]''
\end{itemize}

Refs.~\cite{Tol07C,Hof10,BFB13,Coh17} all conclude that the disturbance is
of second order in the interaction strength using variants of the argument
discussed, and found unconvincing, in Section~\ref{sec:significance}. 
The arguments of Ref.~\cite{DJ12A} are not discussed in this paper,
but in Appendix \ref{app:lindblad} we address some related ideas by the
same authors. 
It is not clear (to this author) what definition is being referred to
in the first quote from Ref.~\cite{Coh17}.

\section{Collective weak measurements}
\label{app:collective}
In the usual protocol for measuring the weak value the variance of $x$ is large 
compared to the shift $\lambda\re\ahat_w$. We thus have to repeat the experiment 
many times to get a good estimate. Another possibility is to let a single meter
interact with a large number $N$ of identically prepared systems.\cite{AV90}
In this case both the uncertainty of the (single) measurement result and the back-action
on each system can, in principle, be made as small as one would like.
 It could thus seem as if this would exclude the
possibility of the kind of statistical effects we have discussed so far. 
An analysis along the lines of Section \ref{sec:significance}, however, shows
that the presence of significant disturbance, if anything, is \emph{more} evident.

We again consider the interaction to be of von Neumann type,
\begin{equation}
  |\psi\rangle^N \rstate \to
  e^{-i\lambda\abar\phat}|\psi\rangle^N \rstate,
  \label{eq:vonN-collective}
\end{equation}
but now $\abar$ is the averaged observable
\begin{equation}
  \abar := N^{-1}\sum_{n=1}^N \ahat_n,
\end{equation}
and we abbreviate
\begin{equation}
  |\psi\rangle^N := |\psi\rangle_1|\psi\rangle_2\cdots|\psi\rangle_N .
\end{equation}
The conditional distribution of $x$ is
\begin{equation}
  P_\lambda(x|\varphi^N,\psi^N) = P_\lambda(\varphi^N|\psi^N)^{-1}\left|\langle x | 
    \left(\langle\varphi|e^{-i\lambda N^{-1}\ahat\phat}|\psi\rangle\right)^{N}\rstate\right|^2 .
\end{equation}
If we now keep $\lambda$ fixed (not necessarily small), but send $N\to\infty$, we can
expand\footnote{We are being somewhat lax in dealing with the unbounded operator $\phat$,
  a more careful treatment can be found in e.g.~Ref.~\cite{AV90}.}
\begin{align}
  \left(\langle\varphi|e^{-i\lambda N^{-1}\ahat\phat}|\psi\rangle\right)^{N}
    &= (\langle\varphi|\psi\rangle)^N[1-i\lambda N^{-1} \ahat_w \phat + O(N^{-2})]^N\nonumber \\
    &= (\langle\varphi|\psi\rangle)^N [e^{-i\lambda\ahat_w\phat} + O(N^{-1})],
\end{align}
leading to
\begin{align}
  P(x|\varphi^N,\psi^N) &= P_\lambda(\varphi^N|\psi^N)^{-1}|\langle\varphi|\psi\rangle|^{2N}
    (2\pi)^{-1/2}\left|e^{-(x-\lambda\ahat_w)^2/4}+ O(N^{-1})\right|^2\nonumber \\
    &= G(x-\lambda\re \ahat_w)+ O(N^{-1}) .
  \label{eq:x-dist-collective}
\end{align}
We again find a shift proportional to the weak value, but now without any condition that
the constant of proportionality $\lambda$ is small.

Turning to the question of back-action, we first note that the strength of the coupling
to each individual system is $N^{-1}\lambda$, which is certainly small in the limit
we are considering.
However, a more relevant quantity in connection with the conditional distribution of $x$ 
is the overall post-selection probability 
\begin{equation}
  P_\lambda(\varphi^N|\psi^N) = |\langle\varphi|\psi\rangle|^{2N}
  	\left[e^{\lambda^2(\im \ahat_w)^2/2}+O(N^{-1})\right] ,
\end{equation}
which is \emph{not} in general close to the unperturbed 
value of $|\langle\varphi|\psi\rangle|^{2N}$.
We see that, when looking at all $N$ systems as a whole, the disturbance 
from the interaction \eqref{eq:vonN-collective} is simply large! It thus seems difficult
to argue that it nevertheless should be considered insignificant.

As for the ordinary weak measurement, it is instructive to work
out the conditional distribution for a measurement of $\xhat' = 2\phat$.
We find 
\begin{align}
  P(x'|\varphi^N,\psi^N) &= P_\lambda(\varphi^N|\psi^N)^{-1}|\langle x' \rstate|^2\left| 
    \langle\varphi|e^{-i\lambda N^{-1}\ahat x'/2}|\psi\rangle\right|^{2N}\nonumber \\
    &= P_\lambda(\varphi^N|\psi^N)^{-1}G(x')|\langle\varphi|\psi\rangle|^{2N}
    	\left[e^{\lambda\im[\ahat_w]x'}+O(N^{-1})\right]\nonumber\\
    &= G(x'-\lambda\im A_w)+ O(N^{-1})
\end{align}
This shift is entirely due to the back-action, but still of the same order of
magnitude as that of $x$ (Eq.~\eqref{eq:x-dist-collective})!

We conclude that, also in the case of collective weak measurements, there is no indication
that the disturbance of the intermediate measurement is negligible.

\section{Disturbance and the Lindblad super-operator}
\label{app:lindblad}
Here we make some remarks on the claims about disturbance in Refs.~\cite{DJ13,Dre14}
(some related ideas can be found in Ref.~\cite{DJ12A}).
We have to begin by making some technical definitions.
First we write the joint probability as
\begin{equation}
  P_\lambda(x,\varphi|\psi) =  
     |\langle\varphi|\mhat_x^\lambda|\psi\rangle|^2,
  \label{eq:joint-p-m}
\end{equation}
where $\mhat_x^\lambda$ is an operator acting on the system Hilbert space. In
the von Neumann case we get the equation
\begin{equation}
  \left|\langle \varphi | \langle x|
    e^{-i\lambda\ahat\phat}|\psi\rangle \rstate\right|^2 = 
     |\langle\varphi|\mhat_x^\lambda|\psi\rangle|^2,
  \label{eq:m-def-eq}
\end{equation}
so we can take $\mhat_x^\lambda$ to be given by (Eq.~\eqref{eq:m-def-eq} does
not uniquely fix $\mhat$, since a phase $\mhat_x\to e^{if_x}\mhat_x$ drops out)
\begin{align}
  \mhat_x^\lambda &= \langle x|e^{-i\lambda\ahat\phat}\rstate \nonumber\\
   &= \langle x|[1-i\lambda\ahat\phat-\frac 1 2 \lambda^2\ahat^2\phat^2+O(\lambda^3)]\rstate 
     \nonumber\\
   &= \sqrt{G(x)}\left[1+\lambda\frac x 2 \ahat + 
   		\frac{\lambda^2}{2}\left(\frac{x^2}{4}-\frac 1 2\right)\ahat^2+O(\lambda^3)\right] .
  \label{eq:m-vonN}
\end{align}
We will not assume that $\mhat_x^\lambda$ is Hermitian, although it happens to be in this
specific example.

The key step of Refs.~\cite{DJ13,Dre14} is to rewrite \eqref{eq:joint-p-m} as
\begin{equation}
  P_\lambda(x,\varphi|\psi) = P^w_\lambda(x,\varphi|\psi) + \mathcal{E}_\lambda(x,\varphi|\psi) ,
\end{equation}
with $P^w$ define by ($\{\cdot,\cdot\}$ denotes the anti-commutator)
\begin{equation}
  P^w_\lambda(x,\varphi|\psi) := 
    |\langle\varphi|\psi\rangle|^2 \re[([\mhat_x^\lambda]^\dagger\mhat_x^\lambda)_w]
   = \frac 1 2 \left\langle 
  \{[\mhat_x^\lambda]^\dagger\mhat_x^\lambda,|\varphi\rangle\langle\varphi|\}\right\rangle_\psi.
\end{equation}
It follows that the `error term' is given by
\begin{equation}
  \mathcal{E}_\lambda(x,\varphi|\psi) = 
    \left\langle \mathcal{L}[\mhat_x^\lambda](|\varphi\rangle\langle\varphi|)\right\rangle_\psi,
  \label{eq:e-def}
\end{equation}
with the Lindblad super-operator\footnote{As is mentioned in both references, 
  Lindblad terms are  association with dissipative 
  dynamics of open quantum systems. It is, however, not clear what the significance 
  of the fact that $\mathcal{E}$ can be written using the Lindblad super-operator is.}
defined as
\begin{equation}
  \mathcal{L}[\mhat](\hat{O}) = 
    \frac 1 2 ([\mhat^\dagger,\hat{O}]\mhat+\mhat^\dagger[\hat{O},\mhat]) .
\end{equation}

It is now observed that, 
if $\mathcal{E}$  can be neglected in the $\lambda\to 0$ limit,
then $P_\lambda(x,\varphi|\psi) \approx P^w_\lambda(x,\varphi|\psi)$ 
and so the conditional expectation value of $x$ is approximately\cite{DJ13,Dre14}
\begin{equation}
  \frac{\int xP^w_\lambda(x,\varphi|\psi)\ud x}{\int P^w_\lambda(x,\varphi|\psi)\ud x} = 
  \frac{|\langle\varphi|\psi\rangle|^2 \int x\re[([\mhat_x^\lambda]^\dagger\mhat_x^\lambda)_w]\ud x}
    {|\langle\varphi|\psi\rangle|^2} = \lambda \re\ahat_w + O(\lambda^2).
\end{equation}
Here we have used (in the von Neumann case there are actually no higher order corrections)
\begin{equation}
 \int x[\mhat_x^\lambda]^\dagger\mhat_x^\lambda \ud x = \lambda\ahat + O(\lambda^2)
 \label{eq:m-obs}
\end{equation}
and
\begin{equation}
  \int [\mhat_x^\lambda]^\dagger\mhat_x^\lambda \ud x = \hat{1}
  \label{eq:m-norm}
\end{equation}
to simplify the numerator and denominator. This argument shows that
\emph{any weak measurement} (i.e.~not necessarily of von Neumann type) that can
be parametrized by operators $\mhat_x^\lambda$ satisfying \eqref{eq:m-obs} and \eqref{eq:m-norm}
will result in $\lambda \re\ahat_w$ as the conditional expectation value, as long
as the error terms can be neglected.\footnote{We have simplified the presentation 
  somewhat compared to Refs.~\cite{DJ13,Dre14}. There Eq.~\eqref{eq:m-obs} is, up to 
  changes in notation, written
  as $\int \alpha_x[\mhat_x^\lambda]^\dagger\mhat_x^\lambda \ud x = \ahat$, where
  the `contextual values' $\alpha_x$ can have complicated dependence on $x$ and $\lambda$.
  The additional generality is not important for the argument we are making.}

The connection with back-action come from the claim\cite{DJ13,Dre14}\footnote{It is 
  expressed most clearly in Ref.~\cite{Dre14} as
  ``[w]hen the error terms $\mathcal E$ are small enough to be neglected [\ldots] (meaning
  that the initial system state is negligibly perturbed), the \emph{real part} of the weak value
  [\ldots] is unambiguously recovered as the \emph{measured} conditioned estimate for $\ahat$,
  verifying our derivation of this real part as a best estimate.''  
} which we will call\\
\textbf{Generalized Disturbance Insignificance (GDI)}: \emph{If the error term 
$\mathcal{E}_\lambda(x,\varphi|\psi)$ can be neglected\footnote{Since we assume
  that we can expand in integer powers of $\lambda$, that 
  $\mathcal{E}_\lambda(x,\varphi|\psi)$ can be neglected means that it is $O(\lambda^2)$.}
then the disturbance caused by the intermediate measurement is insignificant.}

As the names suggests, DI is a special case of GDI. 
To see this, we just need to check that the
error term can be neglected for weak von Neumann measurements. Inserting \eqref{eq:m-vonN}
into \eqref{eq:e-def} we find\footnote{The similarity with \eqref{eq:postselection-prob}
  is no coincidence, since we have the general relation
  $\int\ud x\, \mathcal{E}_\lambda(x,\varphi|\psi) = P_\lambda(\varphi|\psi)
    - |\langle\varphi|\psi\rangle|^2$.}
\begin{equation}
  \mathcal{E}_\lambda(x,\varphi|\psi) = 
    |\langle \varphi |\psi\rangle|^2 G(x)\left[\frac{\lambda^2}{4}
  		\left(|\ahat_w|^2-\re[(\ahat^2)_w]\right)x^2+O(\lambda^3)\right]
\end{equation}
which is indeed $O(\lambda^2)$. This in turn means that the paradoxical consequences of
adopting DI also follow from adopting GDI.
In Ref.~\cite{DJ13} it is remarked that 
\begin{equation}
  P_\lambda(\varphi|\psi)  = \int P_\lambda(x,\varphi|\psi)\ud x
  \approx  \int P^w_\lambda(x,\varphi|\psi) \ud x =  |\langle \varphi|\psi\rangle|^2
\end{equation}
whenever $\mathcal{E}$ can be neglected. This is taken as evidence that
``[t]he Lindblad operation indicates disturbance that the intermediate measurement introduces
to the measurement sequence.''\cite{DJ13} But, as discussed in Section \ref{sec:significance},
the approximate equality $P_\lambda(\varphi|\psi) \approx |\langle \varphi|\psi\rangle|^2$
does not imply that the disturbance is insignificant.

To summarize, GDI is, like DI, an additional postulate which does not follow from ordinary
quantum mechanics. We do not know of any convincing arguments in favor of GDI, while
the arguments against DI works equally well against GDI.

\end{document}